\documentclass{freescale-article}

\journal{freescale}
\pdfoutput=1

\articletype{Research Article}

\usepackage{footnote,upgreek}
\usepackage{multicol}
\usepackage{minitoc}
\usepackage{multirow}

\begin{document}
\title{Freely scalable and reconfigurable optical hardware for deep learning}

\author{Liane Bernstein\authormark{1,$\dagger$,5}, Alexander Sludds\authormark{1,$\dagger$,6}, Ryan Hamerly\authormark{1,2}, Vivienne Sze\authormark{1}, Joel Emer\authormark{3,4}, and Dirk Englund\authormark{1,7}}

\address{\authormark{1}Research Laboratory of Electronics, Massachusetts Institute of Technology, 50 Vassar St, Cambridge, MA 02139, USA\\
\authormark{2}NTT Research Inc.\ Physics \& Informatics Laboratories, 1950 University Ave \#600, East Palo Alto, CA 94303, USA\\
\authormark{3}NVIDIA, 2 Technology Park Drive, Westford, MA 01886, USA\\
\authormark{4}Computer Science and Artificial Intelligence Laboratory, Massachusetts Institute of Technology, 32 Vassar St, Cambridge, MA 02139, USA\\
\authormark{$\dagger$}These authors contributed equally to this work\\
\authormark{5}e-mail: lbern@mit.edu\\
\authormark{6}e-mail: asludds@mit.edu\\
\authormark{7}e-mail: englund@mit.edu}


\begin{abstract*}
As deep neural network (DNN) models grow ever-larger, they can achieve higher accuracy and solve more complex problems. This trend has been enabled by an increase in available compute power; however, efforts to continue to scale electronic processors are impeded by the costs of communication, thermal management, power delivery and clocking. To improve scalability, we propose a digital optical neural network (DONN) with intralayer optical interconnects and reconfigurable input values. The near path-length-independence of optical energy consumption enables information locality between a transmitter and arbitrarily arranged receivers, which allows greater flexibility in architecture design to circumvent scaling limitations. In a proof-of-concept experiment, we demonstrate optical multicast in the classification of 500 MNIST images with a 3-layer, fully-connected network. We also analyze the energy consumption of the DONN and find that optical data transfer is beneficial over electronics when the spacing of computational units is on the order of >10$~\upmu$m. 
\end{abstract*}

\section*{Introduction}

Machine learning has become ubiquitous in modern data analysis, decision-making, and optimization. A prominent subset of machine learning is the artificial deep neural network (DNN), which has revolutionized many fields, including classification~\cite{krizhevsky_imagenet_2012}, translation and prediction~\cite{esteva_dermatologist-level_2017,lecun_deep_2015}. An important step toward unlocking the full potential of DNNs is improving the energy consumption and speed of DNN tasks. To this end, emerging DNN-specific hardware optimizes data access, reuse and communication for mathematical operations: most importantly, general matrix-matrix multiplication (GEMM) and convolution~\cite{sze_efficient_2017}. One approach is to use a specialized memory hierarchy to store and reuse data near an array of computation units, which minimizes reliance on expensive large-scale data distribution networks~\cite{chen_eyeriss:_2017, chen2019eyerissv2, yin_thinker}. Another option for GEMM is a large array of electronic multipliers with fewer intermediate memory tiers~\cite{jouppi_-datacenter_2017}. The large multiplier array reduces overhead, and if the DNN is sizable enough to keep all the processing elements occupied, this design can be more efficient in energy consumption and throughput, thanks to the ability to perform more parallel operations.

\begin{figure}[htbp]
\begin{center}
\includegraphics[width=.95\textwidth]{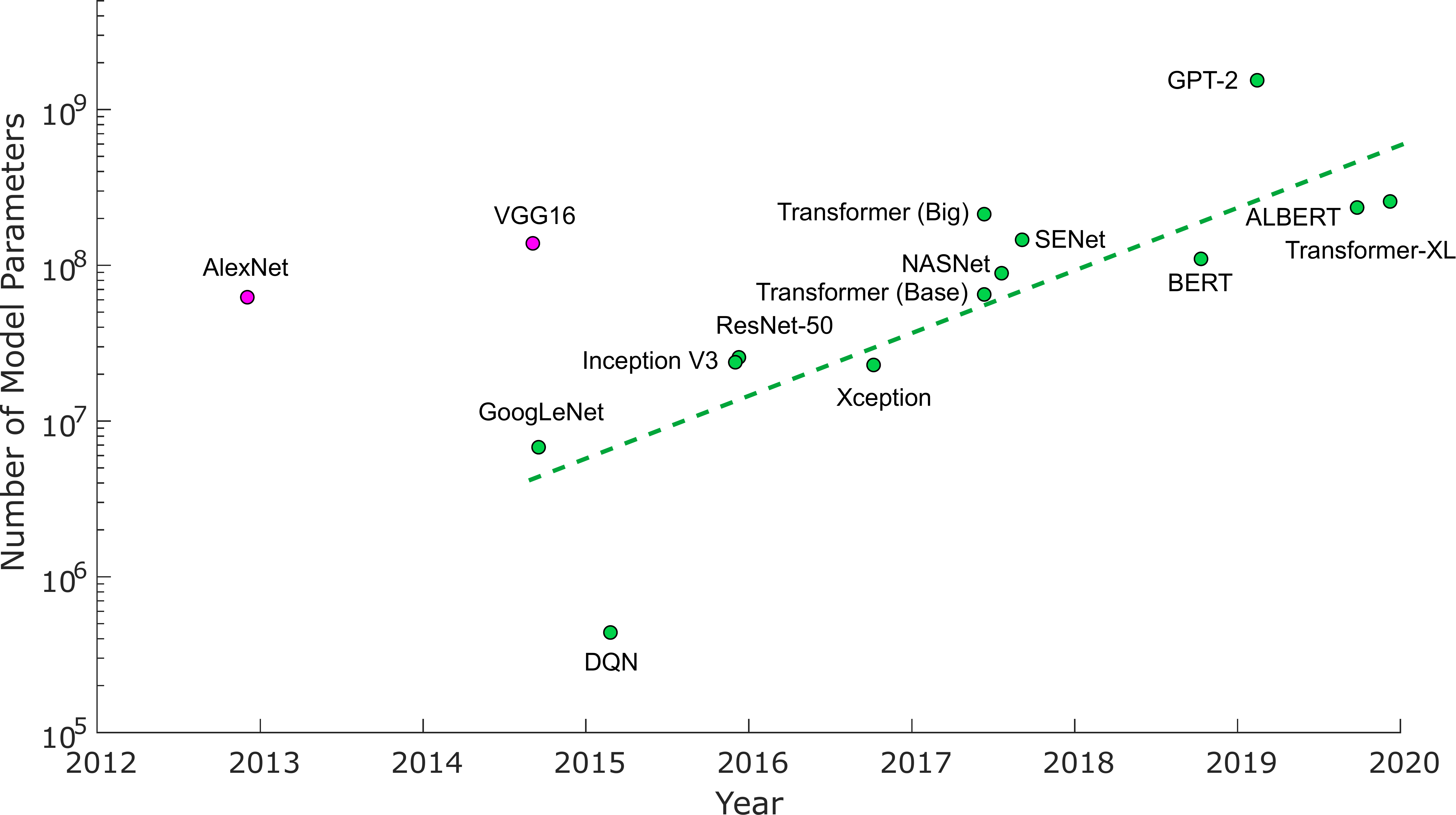}
\caption{Number of parameters, i.e., weights, in recent landmark neural networks~\protect\cite{krizhevsky_imagenet_2012, Simonyan15, szegedy2014going, mnih2015humanlevel, szegedy2016rethinking, heresnet50, chollet2017xception, NIPS2017_7181, nasnet2018, senet2018, devlin2018bert, radford2018gpt2, lan2019albert, dai-etal-2019-transformer} (references dated by first release, e.g., on arXiv). The number of multiplications (not always reported) is not equivalent to the number of parameters, but larger models tend to require more compute power, notably in fully-connected layers. The two outlying nodes (pink) are AlexNet and VGG16, now considered over-parameterized. Subsequently, efforts have been made to reduce DNN sizes, but there remains an exponential growth in model sizes to solve increasingly complex problems with higher accuracy.}
\label{fig:f0}
\end{center}
\end{figure}

However, despite these advances, a central challenge in the field is scaling hardware to keep up with exponentially-growing DNN models (see Fig.~\ref{fig:f0} and Ref.~\cite{xu_scaling_2018}). Many popular DNN models comprise matrices exceeding the GEMM capacity of leading DNN processors (e.g., Google's Tensor Processing Unit (TPU)~\cite{jouppi_-datacenter_2017}), and therefore, matrices must be computed in multiple `tiles'. Tiling requires many inputs and intermediate values to be stored rather than streamed, which increases data movement. Thus, tiling restricts the use of DNNs in high-throughput applications such as the observation of new phenomena in fundamental physics~\cite{duarte_fast_2018, cosmology1, iprijanovi2020deepmerge, neutrino, huerta_enabling_2019}, and reduces the throughput of large DNN models such as recommender systems~\cite{gupta_facebook2020}, vision~\cite{jouppi_-datacenter_2017} and natural language processing~\cite{lan2019albert}. Though the current trend is to scale up conventional electronic hardware, these efforts are impeded by communication~\cite{horowitz_1.1_2014}, clocking~\cite{grs2019}, thermal management~\cite{heat_2012} and power delivery~\cite{gupta_power2007}. Parallel processing with multiple chips~\cite{brainwave} or partitioned chips~\cite{shao_simba:_2019, chiplet_isca} can ease these constraints and improve performance over a monolithic equivalent through greater mapping flexibility~\cite{scalesim}, at the cost of increased communication energy. 

In this Article, we introduce an optical DNN accelerator that encodes data into reconfigurable on-off optical pulses for transmission and passive copying (or \textit{fan-out}) to large-scale electronic multiplier arrays. The near length-independence of optical data routing enables freely scalable systems, where single transmitters are fanned out to many arbitrarily arranged receivers with fast and energy-efficient links. Optics has previously been proposed for analog DNN accelerators, with potential orders-of-magnitude reductions in energy consumption and improved throughput~\cite{hamerly_large-scale_2019, tait_neuromorphic_2017, lin_all-optical_2018, shen_deep_2017,feldmann2020parallel}. In contrast, we propose an entirely digital system, where we replace electrical on-chip interconnects with optical paths for data transmission, but not computation, thus with the capability to preserve accuracy. This `digital optical neural network' (DONN) performs large-scale data distribution from memory to an arbitrary set of electronic multipliers. We first illustrate the DONN architecture and discuss possible implementations. Then, in a proof-of-concept experiment, we demonstrate that digital optical transmission and fan-out with cylindrical lenses has little effect on the classification accuracy of the MNIST handwritten digit dataset (<0.6\%). Crosstalk is the primary cause of this drop in accuracy, and because it is deterministic, it can be compensated: with a simple crosstalk correction scheme, we reduce our bit error rates by two orders of magnitude. Alternatively, crosstalk can be greatly reduced through optimized optical design. Since shot and thermal noise are negligible (see Discussion), the accuracy of the DONN can therefore be equivalent to an all-electronic DNN accelerator.

We also compare the energy consumption of optical interconnects (including light source energy) against that of electronic interconnects over distances representative of logic, memory, and multi-chiplet interconnects in a 7~nm CMOS node. Our calculations show an advantage in data transmission costs for distances $\geq 5$~$\upmu$m (roughly the size of the basic computation unit: an 8-bit multiply-and-accumulate (MAC), with length 5-8~$\upmu$m). Moreover, the DONN scales favorably with respect to very large DNN accelerators that require partitioning into multiple chiplets: the DONN's optical communication cost remains nearly constant at $\sim$0.2~fJ/bit, whereas multi-chiplet systems have much higher electrical interconnect costs ($\sim$90~fJ/bit). Thus, the efficient optical data distribution provided by the DONN architecture will become critical for continued growth of DNN performance through increased model sizes and greater connectivity.

\section*{Results}

\subsection*{Problem statement}

\begin{figure}[htbp]
\begin{center}
\includegraphics[width=1.00\textwidth]{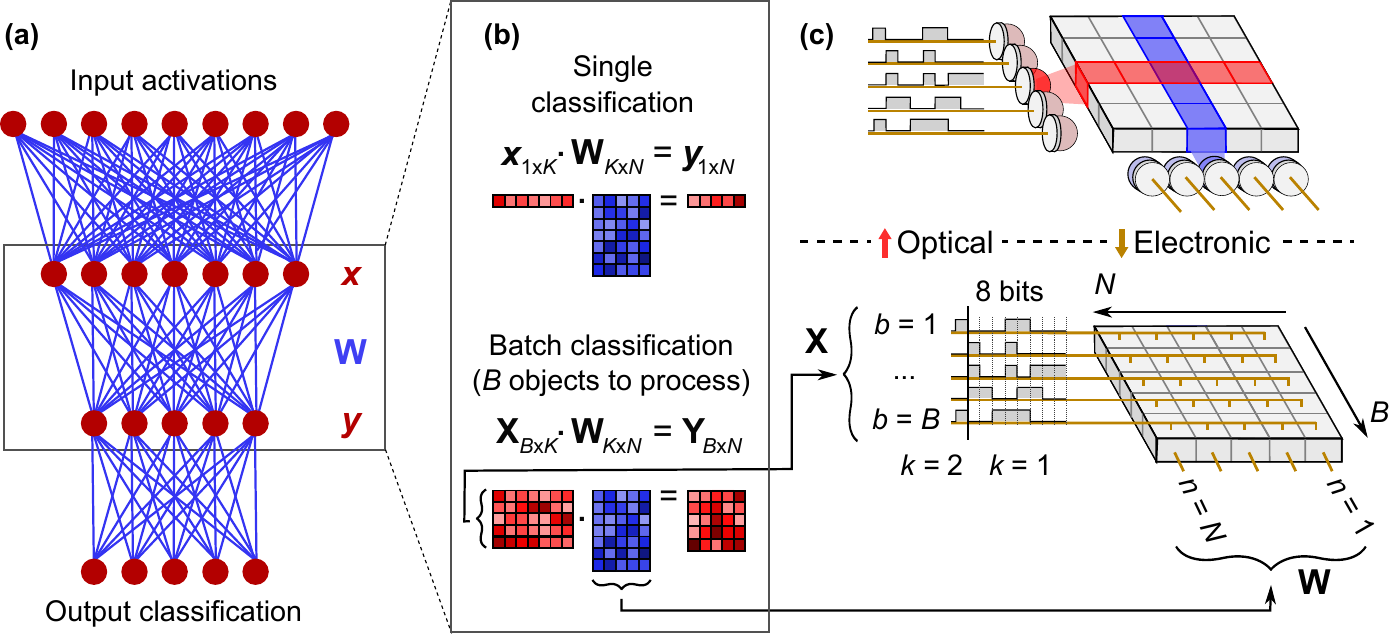}
\caption{Digital fully-connected neural network (FC-NN) and hardware implementations. (a) FC-NN with input activations (red, vector length $K$) connected to output activations (vector length $N$) via weighted paths, i.e., weights (blue, matrix size $K\times N$). (b) Matrix representation of one layer of an FC-NN with $B$-sized batching. (c) Example bit-serial multiplier array, with output-stationary accumulation across $k$. Fan-out of \textbf{X} across $n \in \left\{1...N\right\}$; fan-out of \textbf{W} across $b \in \left\{1...B\right\}$. Bottom panel: all-electronic version with fan-out by copper wire (for clarity, fan-out of \textbf{W} not illustrated). Top panel: digital optical neural network version, where \textbf{X} and \textbf{W} are fanned out passively using optics, and transmitted to an array of photodetectors. Each pixel contains two photodetectors, where the activations and weights can be separated by, e.g., polarization or wavelength filters. Each photodetector pair is directly connected to a multiplier in close proximity.}
\label{fig:f1}
\end{center}
\end{figure}

A DNN consists of a sequence of layers, in which input activations from one layer are connected to the next layer via weighted paths (weights), as shown in Fig.~\ref{fig:f1}a. We focus on inference tasks in this paper (where weights are known from prior training), which, in addition to the energy consumption problem, places stringent requirements on latency and throughput. Modern inference accelerators expend the majority of energy ($>90\%$) on memory access, data movement, and computation in fully-connected (FC) and convolutional (CONV) layers~\cite{chen_eyeriss:_2017}.

Parallelized vector operations, such as matrix-matrix multiplication or successive vector-vector inner products, are the largest energy consumers in CONV and FC layers. In an FC layer, a vector $\boldsymbol{x}$ of input values (`input activations', of length $K$) is multiplied by a matrix \textbf{W}$_{K\times N}$ of weights (Fig.~\ref{fig:f1}b). This matrix-vector product yields a vector of output activations ($\boldsymbol{y}$, of length $N$). Most DNN accelerators process vectors in $B$-sized batches, where the inputs are represented by a matrix \textbf{X}$_{B\times K}$. The FC layer then becomes a matrix-matrix multiplication (\textbf{X}$_{B\times K}\cdot$\textbf{W}$_{K\times N}$). CONV layers can also be processed as matrix multiplications, e.g., with a Toeplitz matrix~\cite{sze_efficient_2017}.

In matrix multiplication, fan-out, where data is read once from main memory (DRAM) and used multiple times, can greatly reduce data movement and memory access. This amortization of read cost across numerous operations is critical for overall efficiency, since retrieving a single matrix element from DRAM requires two to three orders of magnitude more energy than the MAC~\cite{horowitz_1.1_2014}. A simple input-weight product illustrates the benefit of fan-out, since activation and weight elements appear repeatedly, as highlighted by the repetition of $X_{11}$ and $W_{11}$: 
\begin{align}
&\textcolor{white}{W} \begin{bmatrix} \label{eq:matmult}
\textcolor{red}{X_{11}} & X_{12} \\
X_{21} & X_{22}
\end{bmatrix}
\begin{bmatrix}
\textcolor{blue}{W_{11}} & W_{12} \\
W_{21} & W_{22}
\end{bmatrix} 
= \begin{bmatrix}
\textcolor{red}{X_{11}}\textcolor{blue}{W_{11}}+X_{12}W_{21} & \textcolor{red}{X_{11}}W_{12}+X_{12}W_{22}
 \\ X_{21}\textcolor{blue}{W_{11}}+X_{22}W_{21} & X_{21}W_{12}+X_{22}W_{22}
\end{bmatrix}   
\end{align}

Consequently, DNN hardware design focuses on optimizing data transfer and input and weight matrix element reuse. Accelerators based on conventional electronics use efficient memory hierarchies, a large array of tightly packed processing elements (PEs, i.e., multipliers with or without local storage), or some combination of the these approaches. Memory hierarchies optimize temporal data reuse in memory blocks near the PEs to boost performance under the constraint of chip area~\cite{sze_efficient_2017}. This strategy can enable high throughput in CONV layers~\cite{chen_eyeriss:_2017}. With fewer intermediate memory levels, a larger array of PEs (e.g., TPU v1~\cite{jouppi_-datacenter_2017}) can further increase throughput and lower energy consumption on workloads with a high-utilization mapping due to potentially reduced overall memory accesses and a greater number of parallel multipliers (spatial reuse). Therefore, for workloads with large-scale matrix multiplication such as those mentioned in the Introduction, if we maximize the number of available PEs, we can improve efficiency.

\subsection*{Digital optical neural network architecture} \label{sec:architecture}

Our DONN architecture replaces electrical interconnects with optical links to relax the design constraints of reducing inter-multiplier spacing or colocating multipliers with memory. Specifically, optical elements transfer and fan out activation and weight bits to electronic multipliers to reduce communication costs in matrix multiplication, where each element $X_{bk}$ is fanned out $N$ times, and $W_{kn}$ is fanned out $B$ times. The DONN scheme shown in Fig.~\ref{fig:f1}c spatially encodes the first column of \textbf{X}$_{B\times K}$ activations into a column of on-off optical pulses. At the first time step, the activation matrix transmitters fan out the first bit of each of the matrix elements $X_{b1}, \forall b \in \left\{1...B\right\}$ to the PEs (here, $k=1$). Simultaneously, a row of weight matrix light sources transmits the corresponding weight bits $W_{1n}$ to each PE. The photons from these activation and weight bits generate photoelectrons in the detectors, producing the voltages required at the inputs of electronic multipliers (either 0~V for a `0' or 0.8~V for a `1'). After 8 time steps, a multiplier has received $2\times8$~bits (8 bits for the activation value and 8 bits for the weight value), and the electronic multiplication occurs as it would in an all-electronic system. The activation-weight product is completed, and is added to the locally stored partial sum. The entire matrix-matrix product is therefore computed in $8\times K$ time steps; this dataflow is commonly called `output stationary'. Instead of this bit-serial implementation, bits can be encoded spatially, using a bus of parallel transmitters and receivers. The trade-off between added energy and latency in bit-serial multiplication versus increased area from photodetectors for a parallel multiplier can be analyzed for specific applications and CMOS nodes.

\begin{figure}[htbp]
\begin{center}
\includegraphics[width=1\textwidth]{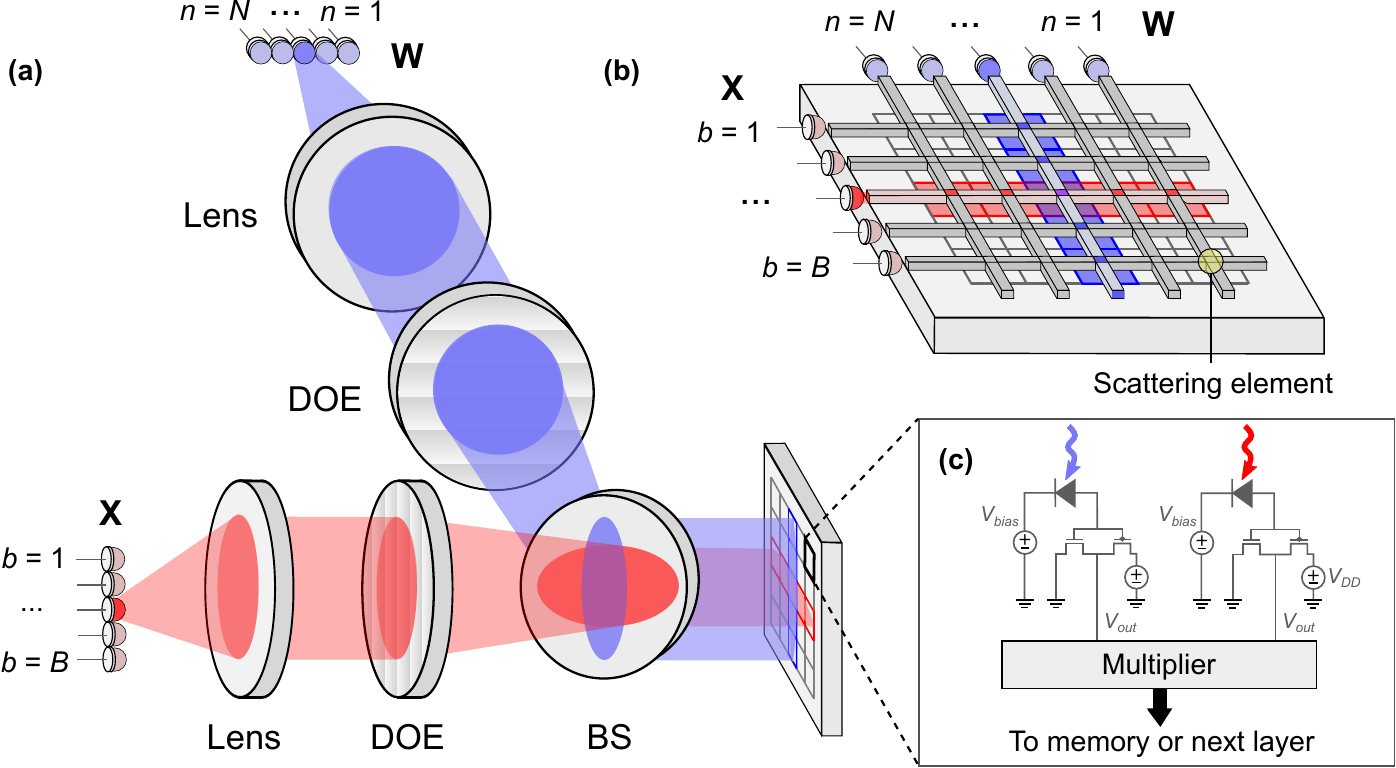}
\caption{Possible implementations of digital optical neural network. (a)~Free-space version. Digital inputs and weights are transmitted electronically to an array of light sources (red and blue, respectively, illustrating different paths). Single-mode light from a source is collimated by a spherical lens (Lens), then focused to a 1D spot array by a diffractive optical element (DOE). A 50:50 beamsplitter brings light from the inputs and weights into close proximity on a custom CMOS receiver. (b)~Waveguide or chip-integrated implementation with scatterers above each processing element (PE). (c)~Example circuit with 2 photodetectors per PE: 1 for activations; 1 for weights. Received bits proceed to multiplier, then memory or next layer.}
\label{fig:f2}
\end{center}
\end{figure}

We illustrate two exemplary experimental DONN implementations in Fig.~\ref{fig:f2}. In the free-space version (Fig.~\ref{fig:f2}a), each source in a linear array of vertical cavity surface emitting lasers (VCSELs) or $\upmu$LEDs emits a cone of light into free space, which is collimated by a spherical lens. A diffractive optical element (DOE) focuses the light to a 1D spot array on a 2D receiver, where the activations and weights are brought into close proximity using a beamsplitter. Figure~\ref{fig:f2}b shows a waveguide or chip-integrated alternative, where each light source is coupled into an optical waveguide. The waveguides are low-loss, except at the scattering elements above each detector pixel. These scattering elements are tuned such that, along one row, an equal amount of light enters each photodetector for a `1' (similar concepts have been experimentally demonstrated~\cite{sun2013large}). In both the free-space and integrated implementations, `receiverless' photodiodes~\cite{miller_attojoule_2017} convert the optical signals to the electrical domain (Fig.~\ref{fig:f2}c). An electronic multiplier then multiplies the values. The output is either saved to memory, or routed directly to another DONN that implements the next layer of computation. Note that the data distribution pattern is not confined to regular rows and columns. A spatial light modulator (SLM), an array of micromirrors, scattering waveguides or a DOE can route and fan out bits to arbitrary locations.

There will be some length-dependent optical loss that will vary based on the implementation. Since free-space propagation is lossless and mirrors, SLMs and diffractive elements are highly efficient (>~95\%), most length- or receiver-number-dependent losses can be attributed to imperfect focusing, e.g., from optical aberrations far from the optical axis. These effects can be mitigated through judicious optical design. There are also waveguide losses in integrated photonics, but these can be very low, e.g., 3~dB/m with mm-scale bend radii in silicon nitride~\cite{bowers_sin2010}. (The DONN does not require any active components, which makes silicon nitride a good choice here.) Therefore, even if we design a meter-length chip with mm-scale bends in the waveguides, we can compensate optical losses by increasing the number of photons generated at the sources (for example, in silicon nitride, by a factor of 2). We assume for the remainder of our analysis that energy is length-independent.

\subsection*{Bit error rate and inference experiments}

\begin{figure}[htbp]
\begin{center}
\includegraphics[width=1.00\textwidth]{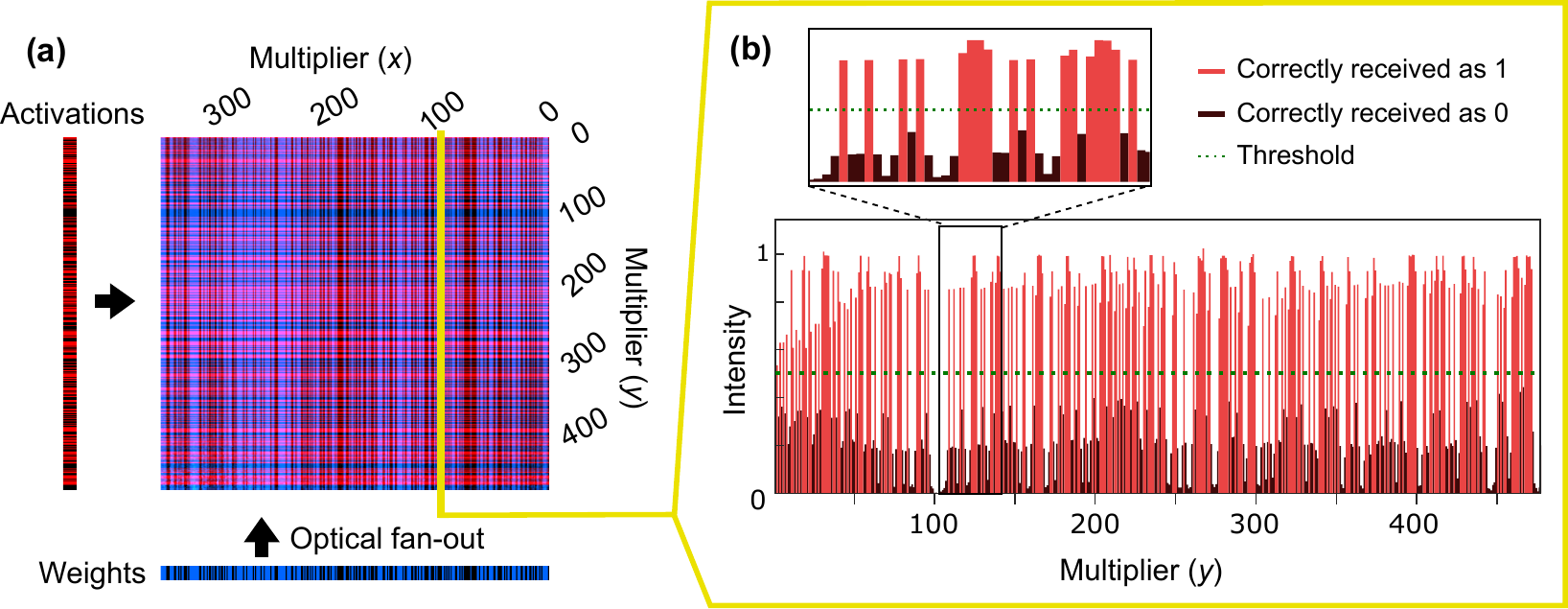}
\caption{Background-subtracted and normalized receiver output from free-space digital optical neural network experiment with random vectors of `1's and `0's diplayed on DMDs. (a) Full 2D image. (b) One column: pixels received as `1' in red and `0' in black.}
\label{fig:f3}
\end{center}
\end{figure}

We used a DONN implementation similar to Fig.~\ref{fig:f2}a to test optical digital data transmission and fan-out for DNNs, as described in Methods. In our first experiment, we determined the bit error rate of our system. Figure~\ref{fig:f3}a shows an example of a background-subtracted and normalized image, captured on the camera when the DMDs displayed random vectors of `1's and `0's. The camera's Bayer filter (described in Methods), as well as optical aberrations and misalignment, caused some crosstalk between pixels (see Fig.~\ref{fig:f3}b). Using a region of $357\times477$ superpixels on the camera, we calculated bit error rates (in a single shot) of $1.2\times 10^{-2}$ and $2.6\times10^{-4}$ for the blue and red channels, respectively. When we confined the region of interest to $151\times191$ superpixels, the bit error rate (averaged over 100 different trials, i.e., 100 pairs of input vectors) was $4.4\times10^{-3}$ and $4.6\times10^{-5}$ for the blue and red arms. See Supplementary Note~1 for more details on bit error rate and error maps. Because crosstalk is deterministic, and not a source of random noise, we can compensate for it. We applied a simple crosstalk correction scheme that assumes uniform crosstalk on the detector and subtracts a fixed fraction of an element's nearest neighbors from the element itself (see Supplementary Note~2). The bit error rates for the blue and red channels then respectively dropped to $2.9\times10^{-3}$ and 0 for the $357\times477$-pixel, single shot image and $2.6\times10^{-5}$ and 0 for the $151\times191$-pixel, 100-image average. In other words, after crosstalk correction, there were no errors in the red channel, and the errors in the blue channel dropped significantly.

\begin{figure}
  \begin{center}
   \includegraphics[width=1.00\textwidth]{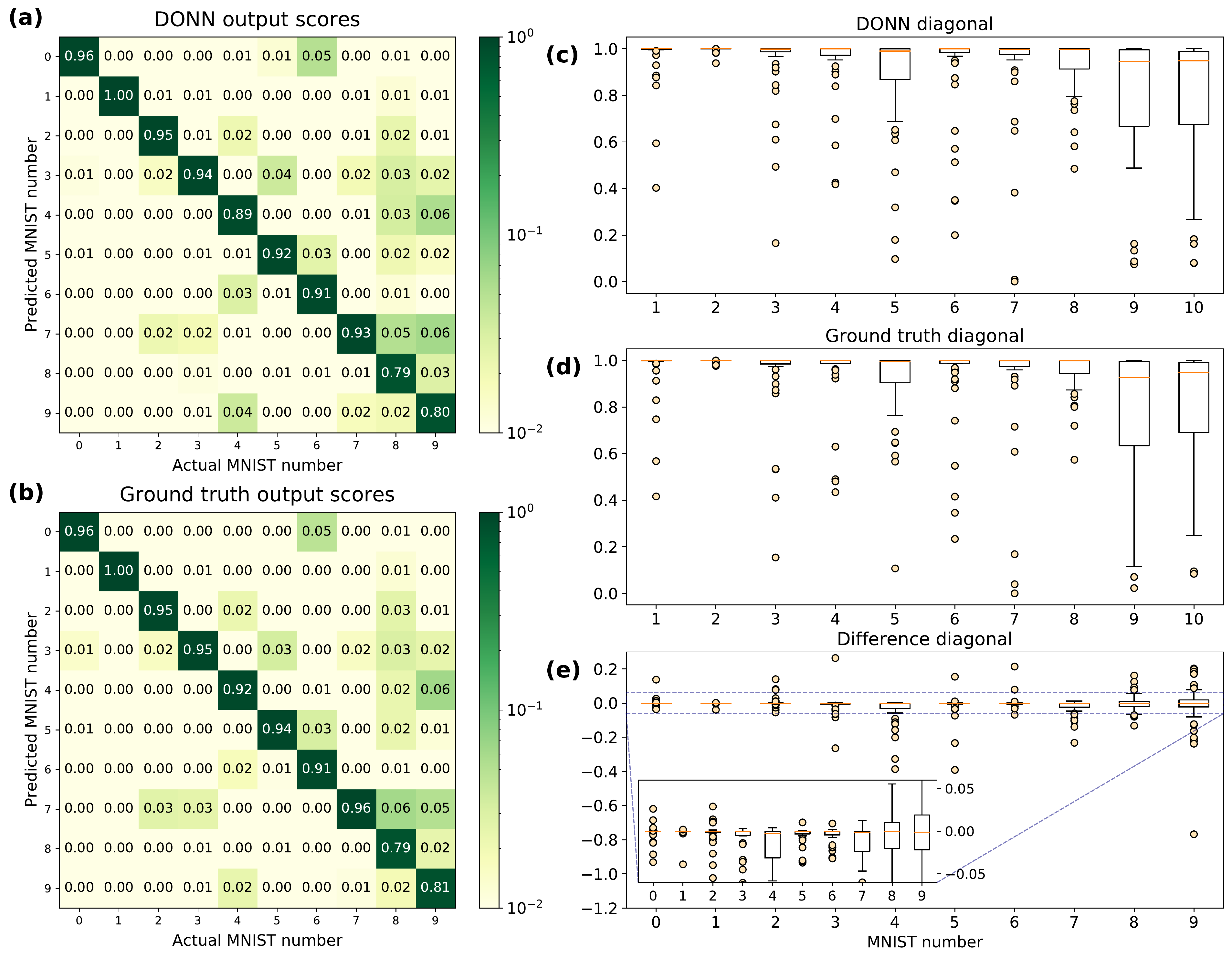}
   \caption{Experimentally measured 3-layer FC-NN output scores (otherwise known as confusion matrix) for 500 MNIST images from test dataset. The values along the diagonal represent correct classification by the model. Each column is an average of $\sim$50 vectors. (a) DONN output scores. (b) Ground-truth (all-electronic) output scores. (c)-(d) Box plot of the diagonals of (a)-(b). (e) Difference in diagonals of DONN versus ground-truth output scores.}
   \label{fig:out_score}
  \end{center}
\end{figure}

Next, we experimentally tested the DONN's effect on the classification accuracy of 500 MNIST images using a three-layer (i.e., two-hidden-layer), fully-connected neural network (FC-NN), with the dataset and training steps described in Supplementary Note~3. We compared our experimental classification results with inference performed entirely on CPU (ground truth) in two ways. The simplest analysis, reported in Table~\ref{tab:t2}, shows a 0.6\% drop in classification accuracy for the DONN versus the ground truth values (or 3 additional incorrectly classified images). Figure~\ref{fig:out_score} illustrates more detailed results, where we analyzed the network output scores. An output score is roughly equivalent to the assigned likelihood that an input image belongs to a given class, and is defined as the normalized (via the softmax function) output vector of a DNN. We found that, along the matrix diagonal, the first and third quartiles in the difference in output scores between the DONN and the ground truth have a magnitude <3\%. The absolute difference in average output scores is also <3\%. We also performed this experiment with a single hidden layer (`2-layer' case), and achieved similar results (a 0.4\% drop in accuracy, or 2 misclassified images). No crosstalk error correction was applied to these results.

\begin{table}[htb]
 \centering \caption{MNIST classification accuracy of DONN versus all-electronic hardware with custom fully-connected neural network models}
\begin{tabular}{ccc}
    \hline
    & 2 layers & 3 layers \\
    \hline
    Electronic (ground truth) & 95.8\% & 96.4\% \\
    DONN & 95.4\% & 95.8\% \\
    \hline
   \end{tabular}
   \label{tab:t2}
\end{table}

\subsection*{Energy analysis: DONN compared with all-electronic hardware} \label{sec:energy}

\begin{figure}[htbp]
\begin{center}
\includegraphics[width=1.00\textwidth]{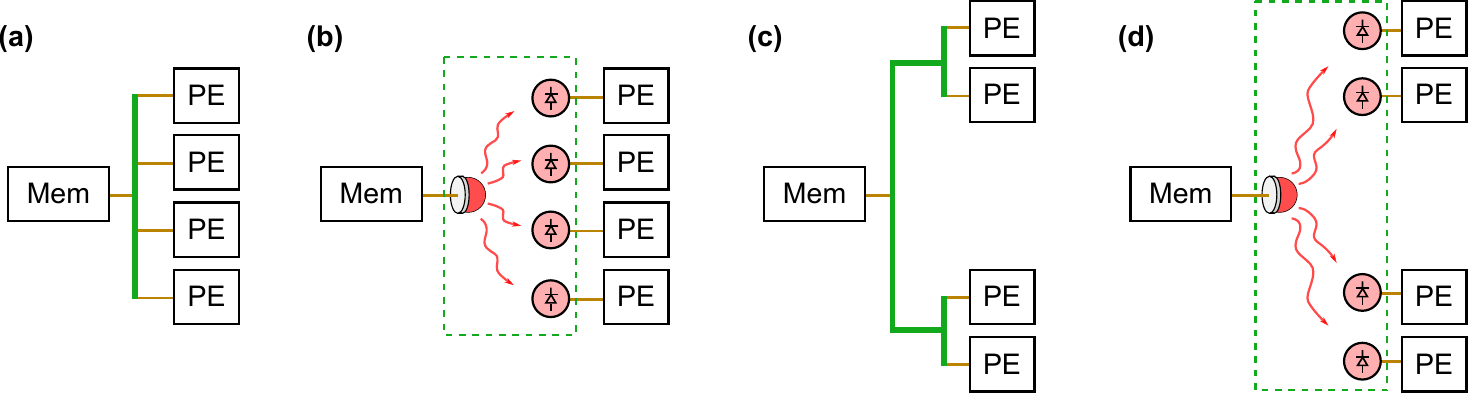}
\caption{Fan-out of one bit from memory (Mem) to multiple processing elements (PEs). (a) Fan-out by electrical wire to a row of PEs in a monolithic chip. (b) DONN equivalent of monolithic chip, where green wire is replaced by optical paths. (c) Fan-out by electrical wire to blocks of PEs divided into chiplets, or separated by memory and logic. (d) DONN equivalent of fan-out to PEs in multiple blocks (energetically equivalent to (b)).}
\label{fig:f4}
\end{center}
\end{figure}

In this section, we compare the theoretical interconnect energy consumption of the DONN with its all-electronic equivalent, where interconnects are illustrated in green in Fig.~\ref{fig:f4}. The interconnect energy, which must include any source inefficiencies, is the energy required to charge the parasitic wire, detector, and inverter capacitances, where a CMOS inverter is representative of the input to a multiplier. See Methods for full energy calculations. In the electronic case, a long wire transports data to a row of multipliers using low-cost (.06~fJ/bit) repeaters (see Supplementary Note~6). The wire has a large parasitic capacitance, but also produces an effective electrical fan-out. In the DONN, the energetic requirements of the detectors contrast with those of conventional optical receivers, which aim to maximize sensitivity to the optical input field, rather than minimize the energetic cost of the system as a whole. The values for electronic and optical components are summarized in Table~\ref{tab:t1}, where $h\nu/e$ must be greater than or equal to the bandgap $E_\text{g}$ of the detector material (here, we have chosen silicon as an example, and set $h\nu/e = E_\text{g}$). $C_\text{det}$ is a theoretical approximation for a $(1\times1\times1)~\upmu\text{m}^3$ cubic detector~\cite{miller_attojoule_2017} and the optical source power conversion efficiency (wall-plug efficiency, i.e., WPE) is a measured value for VCSELs \cite{iga2008vertical,jager199757}. $C_\text{T}$ is an approximation for the capacitance of an inverter in a state-of-the-art node~\cite{zheng_p_finfet_2015, miller_attojoule_2017} and $L_\text{wire}$ is the distance between MAC units in various scenarios.

We find that the optical interconnect energy is independent of length at 0.2~fJ/bit, while the electrical interconnect energy scales from 0.2-0.3~fJ/bit for inter-multiplier communication for abutted MAC units to $90$~fJ/bit for inter-chiplet interconnects. The crossover point where the optical interconnect energy drops below the electrical energy occurs when $L_{\text{wire}} \geq 5~\upmu \text{m}$. The DONN therefore provides an improvement in the interconnect energy for data transmission and can scale to greatly decrease the energy consumption of data distribution with regular distribution patterns. Additionally, advanced technologies are emerging which could lower its energy consumption, such as plasmonic photodetectors with ultra-low capacitance~\cite{tang2008nanometre} and more efficient VCSELs.

\begin{table}[htb]
\footnotesize 
\caption{Interconnect energies over three distances: inter-MAC, inter-SRAM, and inter-chiplet}
    \begin{center}
         \begin{tabular}{|c|c|}
         \multicolumn{2}{c}{Global parameters} \\
    \hline
        $C_{\text{wire}}/\upmu \text{m}$ & $\sim$0.2~fF$/\upmu$m~\cite{miller_attojoule_2017, keckler_gpus_2011, dally_hardwareenabled2018} \\ 
        $C_{\text{T}}$ & $\sim$0.1~fF~\cite{miller_attojoule_2017, zheng_p_finfet_2015} \\ 
        $C_{\text{det}}$ & 0.1~fF~\cite{miller_attojoule_2017} \\ 
        $h \nu / e$ & 1.12~\text{eV}\\
        WPE & $\sim$0.5~\cite{iga2008vertical,jager199757}\\ 
    \hline
  \end{tabular}
    \end{center}
   
  \vfill
    \begin{tabular}{|c|c|}
    \multicolumn{2}{c}{Inter-MAC (8-bit MAC unit)} \\
    \hline
        $L_\text{wire}$ & 5-8~$\upmu\text{m}^\dagger$ \\
        $V_{DD}$ & 0.80~V~\cite{stillmaker2017scaling} \\ 
        $E_\text{elec}/\text{bit}$ & 0.2-0.3~fJ/bit \\ 
        $E_\text{DONN}/\text{bit}$ & 0.2~fJ/bit \\ 
    \hline
  \end{tabular}
    \hfill
    \begin{tabular}{|c|c|}
     \multicolumn{2}{c}{Inter-SRAM (7~nm SRAM macro)} \\
    \hline
        $L_\text{wire}$ & 60~$\upmu$m ~\cite{chang201712} \\
        $V_{DD}$ & 0.75~V*~\cite{chang201712} \\ 
        $E_\text{elec}/\text{bit}$ & 2~fJ/bit \\ 
        $E_\text{DONN}/\text{bit}$ & 0.2~fJ/bit \\ 
    \hline
  \end{tabular}
    \hfill
    \begin{tabular}{|c|c|}
    \multicolumn{2}{c}{Inter-chiplet~\cite{shao_simba:_2019}} \\
    \hline
        $L_\text{wire}$ & $\sim$2500~$\upmu$m \\
        $V_{DD}$ & 0.85~V*\\ 
            $E_\text{elec}/\text{bit}$ & $90$~fJ/bit\\ 
        $E_\text{DONN}/\text{bit}$ & 0.2~fJ/bit \\ 
    \hline
  \end{tabular}
  \caption*{$^\dagger$We assume a square multiplier and scale reported 8-bit multiplier areas~\cite{saadat_minimally_2018, shoba_energy_2017, ravi_design_2015} from a 45~nm to a 7~nm node (the current state of the art) with the scaling factors from Ref.~\cite{stillmaker2017scaling}. A MAC unit comprises both an 8-bit multiplier and a 32-bit adder, so we are placing a lower bound on the minimum length of $L_\text{wire}$. Recent work~\cite{johnson_mac} optimizes MAC units for DNNs, and reports a $337~\upmu\text{m}^2$ area in a 28~nm node, where the MAC unit comprises an 8-bit multiplier and a 32-bit adder. Extrapolated to a 7~nm node with a fourth-order polynomial fit of the scaling factors from Ref.~\cite{stillmaker2017scaling}, the MAC unit is of size ($7~\upmu\text{m})^2$, which falls within the 5-8~$\upmu$m range.}
  \caption*{*Input-output voltage and core logic voltage can differ in CMOS. In optics, however, since the data delivery mechanism does not vary with distance travelled, we assume $V_{DD}$ remains constant at 0.80~V.}
    \label{tab:t1}
\end{table}

\section*{Discussion} \label{sec:discussion}

With minimal impact on accuracy, the DONN yields an energy advantage over all-electronic accelerators with long wire lengths. In our proof-of-concept experiment, we performed inference on 500 MNIST images with 2- and 3-layer FC-NNs and found a <0.6\% drop in accuracy and a <3\% absolute difference in average output scores with respect to the ground truth implementation on CPU. We attributed these errors to crosstalk due to imperfect alignment and blurring from the camera's Bayer filter. In fact, a simple crosstalk correction scheme lowered measured bit error rates by two orders of magnitude. We could thus transmit bits with 100\% measured fidelity in the activation arm (better aligned than the weight arm), which illustrates that crosstalk can be mitigated and possibly eliminated either through post-processing, charge sharing at the transmitters, greater spacing of receivers, or optimized design of optical elements and receiver pixels. In the hypothetical regime where error due to crosstalk is negligible, the remaining noise sources are shot and thermal noise. Intuitively, shot and thermal noise are also present in an all-electronic system, and the number of photoelectrons at the input to an inverter in the DONN is equal to the number of electrons at the input to an inverter in electronics. Therefore, if these noise sources do not limit accuracy in the all-electronic case, the same can be said for the DONN~\cite{miller_attojoule_2017}. For mathematical validation that shot and thermal noise have a trivial impact on bit error rate in the DONN, see Supplementary Note~7. These analyses demonstrate that the fundamental limit to the accuracy of the DONN is no different than the accuracy of electronics, and thus, we do not expect accuracy to hinder DONN scaling in an optimized system.

In our theoretical energy calculations, we compared the nearly length-independent data delivery costs of the DONN with those of an all-electronic system. We found that in the worst case, when multipliers are abutted in a multiplier array, optical transmitters have a similar interconnect energy cost compared to copper wires in a 7~nm node ($\sim$0.2~fJ/bit versus $\sim$0.2-0.3~fJ/bit). The regime where the DONN shows important gains over copper interconnects is in architectures with increased spacing between computation units, e.g., with locally-packed memory and logic ($\sim$0.2~fJ/bit versus $\sim$2~fJ/bit), or with multiple chiplets ($\sim$0.2~fJ/bit versus $\sim$90~fJ/bit). In the multi-chiplet case, the cost to transmit two 8-bit values in electronics ($\sim$1,400~fJ) is therefore significantly larger than that of an 8-bit MAC (25~fJ)~\cite{stillmaker2017scaling,horowitz_1.1_2014}. On the other hand, in optics, the interconnect cost ($\sim$3~fJ for 2$\times$8~bits, including source energy) remains an order of magnitude smaller than the MAC cost. Since multi-chiplet and multi-chip systems offer a promising approach to increasing throughput on large DNN models, optical connectivity can further these scaling efforts by reducing inter-chiplet communication energy by orders of magnitude.

In addition, because length-independent data distribution is a tool currently unavailable to digital system designers, relaxing electronic constraints on locality can open new avenues for DNN accelerator architectures. For example, memory can be devised such that numerous small pieces of memory are located far away from the point of computation and reused many times spatially, with a small fixed cost for doing so. Designers can then lay out smaller memory blocks with higher bandwidth, lower energy consumption, and higher yield. If we keep memory and computation spatially distinct, we have the added benefit of allowing for more compact memories that consume less energy and area, e.g., DRAM, which is fabricated with a different process than typical CMOS to achieve higher density than on-chip memories. Furthermore, due to its massive fan-out potential, the DONN can, firstly, reduce overhead by minimizing a system's reliance on a memory hierarchy and, secondly, amortize the cost of weight delivery to multiple clients running the same neural network inference on different inputs. Additionally, some newer neural network models require irregular connectivity (e.g., graph neural networks, which show state-of-the-art performance on recommender systems, but are restricted in size due to insufficient compute power~\cite{graph_survey_wu2020, graph_survey_zhang2020}). These systems have arbitrary connections with potentially long wire lengths between MAC units, representing different edges in the graph. The DONN can implement these links without incurring additional costs in energy from a complex network-on-chip in electronics. Yet another instance of greater distance between multipliers is in higher-bit-precision applications, as in training, which require larger MAC units. Lastly, the DONN could facilitate thermal management in chips, with the option to increase spacing between compute units at no extra cost.

In future work, we plan to assess the performance of the DONN on state-of-the-art DNN workloads, such as the models described in MLPerf~\cite{mlperf_micro2020}. Firstly, we will benchmark the DONN against all-electronic state-of-the-art accelerators by using Timeloop~\cite{parashar_timeloop:_2019}. Through a search for optimal mappings (ways to organize data and computation), this software can simulate the total energy consumption and latency of running various workloads on a given hardware architecture, including computation and memory access. Timeloop therefore enables us to perform an in-depth comparison of all-electronic accelerators against the proposed instances of the DONN, including variable data transmission costs for different electronic wire lengths, and waveguide losses in the chip-integrated DONN. Second, we will design an optical setup and receiver to reduce experimental crosstalk, power consumption and latency. We can then test larger workloads on this optimized hardware. Finally, beyond neural networks, there are many examples of matrix multiplication which a DONN-style architecture can accelerate, such as optimization, Ising machines and statistical analysis, and we plan to investigate these applications as well.

In summary, the DONN implements arbitrary transmission and fan-out of data with an energy cost per bit that is nearly independent of data transmission length and number of receivers. This property is key to scaling deep neural network accelerators, where increasing the number of processing elements for greater throughput in all-electronic hardware typically implies higher data communication costs due to longer electronic path length. Contrary to other proposed optical neural networks~\cite{hamerly_large-scale_2019, tait_neuromorphic_2017, lin_all-optical_2018, shen_deep_2017,feldmann2020parallel}, the DONN does not require digital-to-analog conversion and is therefore less prone to error propagation. The DONN is also reconfigurable, in that the weights and activations can be easily updated. Our work indicates that the nearly length-independent communication enabled by optics is useful for digital neural network system design, for example to simplify memory access to weight data. We find that optical data transfer begins to save energy when the spacing of MAC computational units is on the order of $>$10~$\upmu$m. More broadly, further gains can be expected through the relaxation of electronic system architecture constraints.

\section*{Methods}

\subsection*{Digital optical neural network implementation for bit error rate and inference experiments}

We performed bit error rate and inference experiments with optical data transfer and fan-out of point sources using cylindrical lenses. Two digital micromirror devices (DMDs, Texas Instruments DLP3000, DLP4500) illuminated by spatially-filtered and collimated LEDs (Thorlabs M625L3, M455L3) acted as stand-ins for the two linear source arrays. For the input activations/weights, each 10.8~$\upmu$m-long mirror in one DMD column/row either reflected the red/blue light toward the detector (`1') or a beam dump (`0'). Then, for each of the DMDs, an $f=100~\text{mm}$ spherical lens followed by an $f=100~\text{mm}$ cylindrical achromatic lens imaged one DMD pixel to an entire row/column of superpixels of a color camera (Thorlabs DCC3240C). Each camera superpixel is made up of four pixels of size (5.3~$\upmu$m)$^2$: two green, one red and one blue. The camera acquisition program applies a `de-Bayering' filter to automatically extract color information for each sub-pixel; this filter causes blurring, and therefore it increased crosstalk in our system. In a future version of the DONN, a specialized receiver will reduce this crosstalk and also operate at a higher speed.

To process the image received on the camera, we subtracted the background, normalized, then thresholded. (We acquired normalization and background curves with all DMD pixels in the `on' and `off' states, respectively. This background subtraction and normalization could be implemented on-chip by precharacterizing the system, and biasing each receiver pixel by some fixed voltage.) If the detected intensity was above the threshold value, it was labeled a `1'; below threshold, a `0'. For the bit error rate experiments, we compared the parsed values from the camera with the known values transmitted by the DMDs, and defined the bit error rate as the number of incorrectly received bits divided by the total number of bits. In the inference experiments, the DMDs displayed the activations and pre-trained weights, which propagated through the optical system to the camera. After background subtraction and normalization, the CPU multiplied each activation with each weight, and applied the nonlinear function (ReLU after the hidden layers and softmax at the output). We did not correct for crosstalk here, to illustrate the worst-case scenario of impact on accuracy. The CPU then fed the outputs back to the input activation DMD for the next layer of computation. We used a DNN model with two hidden layers with 100 activations each and a 10-activation output layer. We also tested a model with a single hidden layer with 100 activations.

\subsection*{MNIST preprocessing}
For the inputs to the network, a bilinear interpolation algorithm transformed the $28\times28$-pixel images into $7\times7$-pixel images, which were then flattened into a 1D 49-element vector. The following standard mapping quantized both input and weight matrices into 8-bit integer representations:
\begin{align}
\rm{Quantized} = \rm{QuantizedMin} + \frac{(\rm{Input} - \rm{Floating Min})}{\rm{Scale}}    
\end{align}

\noindent where Quantized is the returned value, QuantizedMin is the minimum value expressible in the quantized datatype (here, always 0), Input is the input data to be quantized, FloatingMin is the minimum value in Input, and Scale is the scaling factor to map between the two datatype ranges $\left(\frac{\rm{FloatingMax} - \rm{FloatingMin}}{\rm{QuantizedMax} - \rm{QuantizedMin}}\right)$. See gemmlowp documentation \cite{gemmlowp} for more information on implementations of this quantization. (In practice, 8-bit representations are widely used in DNNs, since 8-bit MACs are generally sufficient to maintain accuracy in inference~\cite{judd2016stripes, albericio2017bitpragrmatic, jouppi_-datacenter_2017}).

\subsection*{Electronic and optical interconnect energy calculations}
When an electronic wire transports data over a distance $L_\text{wire}$ to the gate of a CMOS inverter (representative of a full-adder's input, which are the building blocks of multipliers), the energy consumption per bit is:
\begin{align} \label{eq:electronic}
    E_\text{elec}/\text{bit} = \tfrac{1}{4}\left(\tfrac{C_{\text{wire}}}{\upmu \text{m}} \cdot L_{\text{wire}}+C_\text{T}\right) \cdot V_{DD}^2
\end{align}
where $V_{DD}$ is the supply voltage, $L_{\text{wire}}$ is the wire length between two multipliers and $C_\text{T}$ is the inverter capacitance. Interconnects consume energy predominantly when a load capacitance, such as a wire, is charged from a low (0~V) to a high ($\sim$1~V) voltage, i.e., in a $0\rightarrow1$ transition. If we assume a low leakage current, maintaining a value of `1' (i.e., $1\rightarrow1$) consumes little additional energy. To switch a wire from a `1' to a `0', the wire is discharged to the ground for free (Supplementary Note~4). Lastly, maintaining a value of `0' simply keeps the voltage at 0~V, at no cost. Assuming a random distribution of `0' and `1' bits, we therefore include a factor of 1/4 in equation~(\ref{eq:electronic}) to account for this dependence on switching activity.

In the DONN, a light source replaces the wire for fan-out. The low capacitances of the receiverless detectors in the DONN allow for the removal of receiving amplifiers~\cite{miller_attojoule_2017}. Thus, the DONN's minimum energy consumption corresponds to the optical energy required to generate a voltage swing of 0.8~V on the load capacitance (i.e., the photodetector ($C_\text{det}$) and an inverter ($C_\text{T}$)), all divided by the source's power conversion efficiency (called wall-plug efficiency, WPE). Subsequent transistors in the multiplier are powered by the off-chip voltage supply, as in the all-electronic architecture. Assuming a detector responsivity of $\sim$1~\cite{miller2013responsivity}, the DONN interconnect energy cost is:
\begin{align} \label{eq:optical}
    E_\text{DONN}/\text{bit} = \tfrac{1}{2\cdot \text{WPE}}\cdot h\nu \cdot n_\text{p}
\end{align}
\noindent where $h\nu$ is the photon energy and the number of photons per bit, $n_\text{p}$, is determined by:
\begin{align} \label{eq:num_phot}
    n_\text{p} =\frac{\left(C_{\text{det}}+C_\text{T}\right) \cdot V_{DD}}{e}
\end{align}
\noindent As in the all-electronic case, we assume low leakage on the receiverless photodetector. Photons are received for every `1' and therefore, to avoid charge buildup, charge on the output capacitor must be reset after every clock cycle. In Supplementary Note~5, we propose a CMOS discharge circuit that actively resets the receiver. (Another possible method is a dual-rail encoding scheme \cite{miller_attojoule_2017}.) Thus, the switching activity factor is 1/2 instead of 1/4: as for the all-electronic case, we assume a random distribution of bits, but here, both $1\rightarrow1$ and $0\rightarrow1$ have a nonzero cost.


\bibliography{ms}

\section*{Acknowledgements}

Thanks to Christopher Panuski for helpful discussions about $\upmu$LEDs and Angshuman Parashar and Yannan (Nellie) Wu for insights into all-electronic DNN accelerators. We would also like to thank Mohamed Ibrahim for useful discussions on receiver discharging circuits. Anthony Pennes helped with several machining tasks. Thanks to Ronald Davis III and Zhen Guo for manuscript revisions. We also thank the NVIDIA Corporation for the donation of the Tesla K40 GPU used for training the fully-connected networks. Equipment was purchased thanks to the U.S. Army Research Office through the Institute for Soldier Nanotechnologies (ISN) at MIT under grant no. W911NF-18-2-0048. L.B. is supported by a Postgraduate Scholarship from the Natural Sciences and Engineering Research Council of Canada, National Science Foundation (NSF) E2CDA grant no. 1640012 and the afore-mentioned ISN grant. A.S. is supported by an NSF Graduate Research Fellowship Program under grant no. 1122374, NTT Research Inc., NSF EAGER program grant no. 1946967, and the NSF/SRC E2CDA and ISN grants mentioned above. R.H. was supported by an Intelligence Community Postdoctoral Research Fellowship at MIT, administered by ORISE through the U.S. DoE / ODNI.

\section*{Author contributions}
D.E. and R.H. developed the original concept. L.B. designed and performed the hardware experiments with the support of A.S. and D.E. A.S. developed the data acquisition, training, and confusion matrix analysis software. L.B. developed the output image processing software and performed the bit error rate calculations. L.B. and A.S. performed the energy calculations, with critical insights from R.H. J.E. and V.S. provided critical insights into all-electronic hardware comparisons. L.B. and A.S. wrote the manuscript with input from all authors. R.H., J.E., V.S. and D.E. supervised the project.

\section*{Competing interests}
The authors declare no competing interests.

\section*{Additional information}

\subsection*{Correspondence}
Correspondence and requests for materials should be addressed to L.B., A.S. or D.E.

\subsection*{Supplementary information}
Supplementary information is available for this paper.

\end{document}


\maketitle

\section*{Supplementary Note 1: Bit error rate due to crosstalk}
\begin{figure}
    \centering
    \includegraphics[scale=.8]{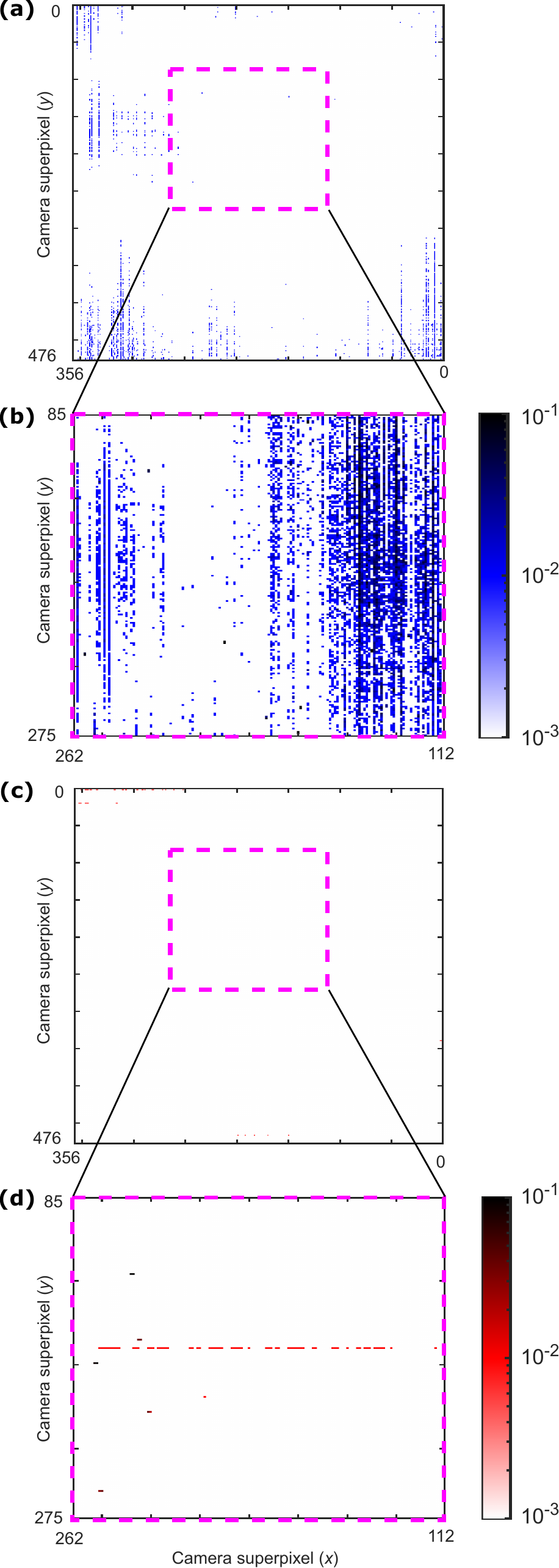}
    \caption{Bit error rates in proof-of-concept experiment. (a) Blue channel: errors when random vector of `0's and `1's displayed on DMD (single shot). Blue: incorrectly transmitted bit; white: correct. (b) Region of interest selected for experiment. Error in blue channel averaged over 100 frames (different vectors displayed on DMD at each frame). (c)-(d) same as (a)-(b), but for red channel.}
    \label{fig:error}
\end{figure}

Here, we show experimental bit error rate maps for both the blue and red channels. Each DMD pixel is fanned out to a row (column) of superpixels on the camera for the input activations (weights). The Bayer filter allows the discrimination of the input activations from the weights into red and blue channels, respectively. Since the camera has four sub-pixels per superpixel, we bin the sub-pixels into $2\times2$ blocks. As described and shown in Fig.~4 of the main text, random vectors of `1's and `0's were displayed on the DMDs to assess bit error rates in data transmission from two 1D source arrays to the camera. In Fig.~\ref{fig:error}, we show bit error rate maps. Images~\ref{fig:error}(a) and (c) show the error from a single shot (one random vector pair displayed on the DMDs). Images~\ref{fig:error} (b) and (d) show the error averaged over 100 frames (100 different random vector pairs displayed on the DMDs) in the low-error region of interest used for the proof-of-concept experiment. The error is larger on the edges of each image due to optical aberrations, and is larger in the blue channel than the red channel due to misalignment.

\section*{Supplementary Note 2: Crosstalk correction}
The bit error rate described in the previous section is mainly attributable to optical crosstalk at the detector, due to imperfect lenses and alignment. Since this error is deterministic (as opposed to random fluctuations), it can be compensated by post-processing. To illustrate this principle, we performed simple crosstalk correction: we multiplied each line of an image detected on the camera by a tridiagonal crosstalk reduction matrix, per equation~(\ref{eq:xtalk}) (where  $\overline{I_{:n}}$ is the corrected line of the camera image).

\begin{figure}[htbp]
    \centering
    \includegraphics[scale=.9]{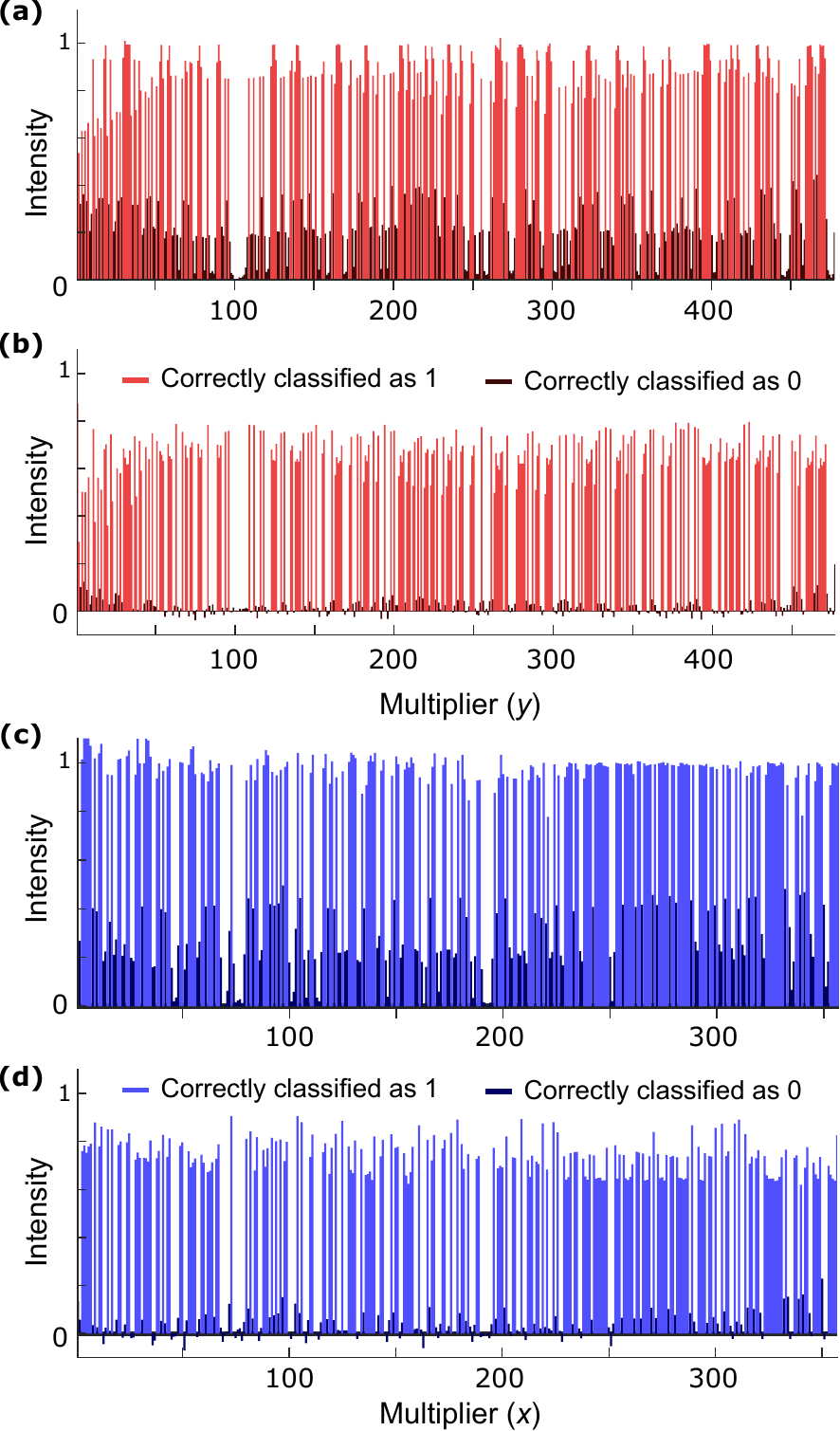}
    \caption{One line of receiver image after background subtraction and normalization, with random vectors of `1's and `0's displayed on DMDs. (a) Column 100 in red channel (same as Fig. 4b in main text). (b) Same as (a), after crosstalk correction. (c) Row 100 in blue channel. (d) Same as (c), after crosstalk correction.}
    \label{fig:xtalk_corr}
\end{figure}

$\xi$ was estimated to be $\sim$0.19 and $\sim$0.18 for the red and blue arms, respectively, from a calibration image of alternating `1's and `0's transmitted by the DMDs. $\overline{I_{:n}}$ is renormalized after this matrix multiplication. We show the effects of crosstalk reduction in Fig.~\ref{fig:xtalk_corr}.

\begin{align}
\begin{bmatrix}
\overline{I_{1n}} \\ \overline{I_{2n}} \\ \vdots \\ \\ \\
\end{bmatrix}
=
\begin{bmatrix}
1 & -\xi & 0 & & \\
-\xi & 1 & -\xi & & \\
0 & -\xi & 1 & \ddots & \\
& & \ddots & \ddots & \\
& & & & 1
\end{bmatrix}
\begin{bmatrix}
I_{1n} \\ I_{2n} \\ \vdots \\ \\ \\
\end{bmatrix} \label{eq:xtalk}
\end{align}

To maximize energy efficiency and throughput, the final version of this system (with a custom CMOS chip that integrates detection with digital MAC computation) will not perform any post-processing. Instead, we might use a charge-sharing scheme at the transmitters to implement a version of equation~(\ref{eq:xtalk}). Alternatively, we could simply reduce crosstalk by changing the system design; for example, we could choose to space the PEs further apart or shrink the active region of the detectors to improve the ratio of signal at the current pixel to noise from neighboring pixels.

\section*{Supplementary Note 3: Training and test sets}
In our proof-of-concept experiment, we performed inference on 500 images using a two-hidden-layer, fully-connected neural network, where each hidden layer had 100 activations. We used TensorFlow's built-in dataset importer to download the first 500 images in the test set of the MNIST handwritten digit dataset \cite{lecun-mnist}, as downloaded from the TensorFlow~2 Keras database. Relevent code can be found in the GitHub repository for user Alexander Sludds (alexsludds):

\noindent\url{https://github.com/alexsludds/Digital-Optical-Neural-Network-Code}

The model's weights were pre-trained on an NVIDIA K40 GPU using the entire MNIST training set. Categorical cross-entropy was used as a loss function. Dropout regularized the model's weights in each layer to prevent overfitting. Input images were downsized from $49\times49$ to $7\times7$ using bilinear interpolation.

\section*{Supplementary Note 4: Electronic interconnect switching energy in 0 to 1 transitions}

\begin{figure}
    \centering
    \includegraphics[scale=0.4]{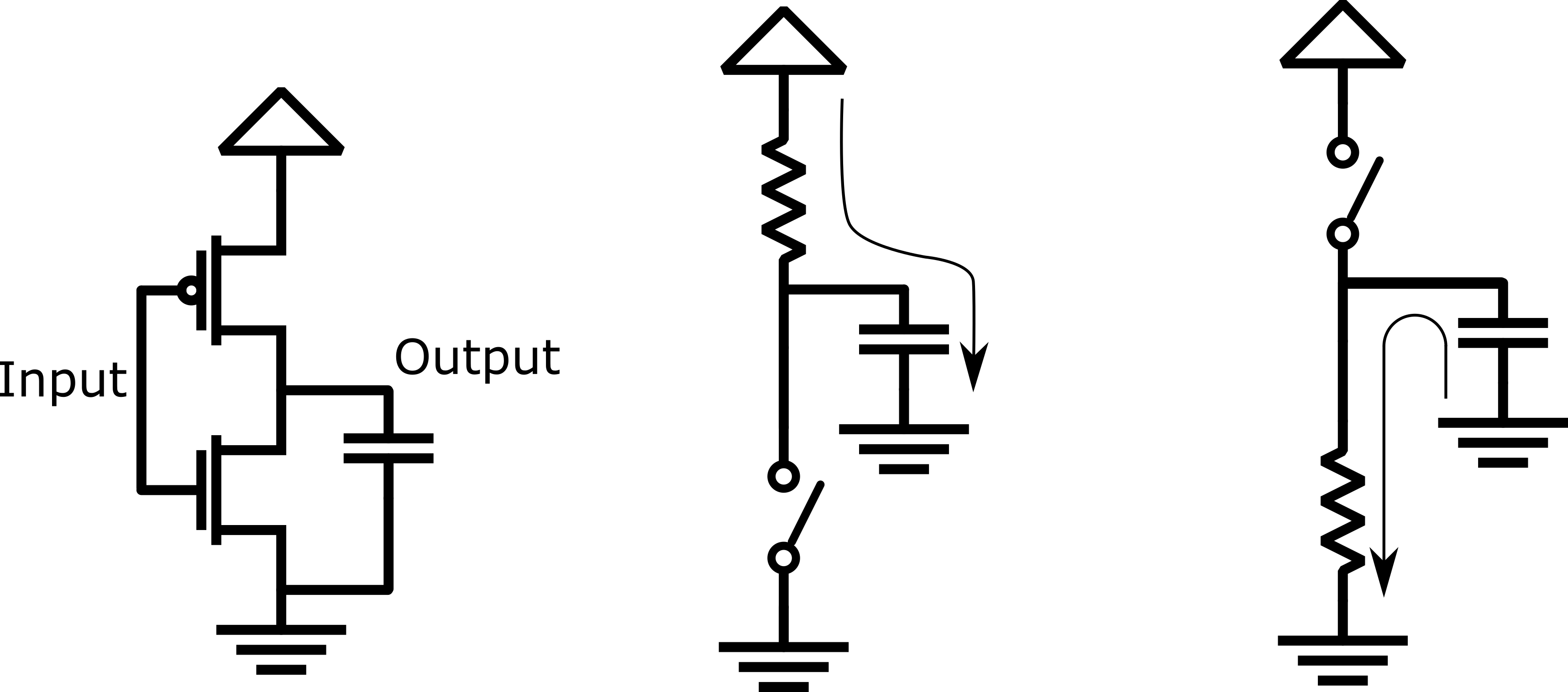}
    \caption{A demonstration of where dynamic energy consumption goes during switching of a CMOS inverter. The circuit, shown left, consists of a stacked NMOS and PMOS device. During an output low to high transition, shown center, charge is deposited on the lumped output capacitance. During an output low to high transition, shown right, that charge from the lumped output capacitance is discharged through the NMOS into ground.}
    \label{fig:switching}
\end{figure}

The dynamic switching energy of CMOS devices is the amount of energy required to charge the output capacitance of a CMOS gate. Energy is only consumed in CMOS inverters for low-to-high transitions on the outputs of these gates. Consider the toy circuit model shown in Fig. \ref{fig:switching}. On the left is a CMOS inverter, and on the right are a low-to-high and high-to-low transitions, respectively. In the low-to-high transition, the PMOS has to switch closed, shorting the output to the supply rail by charging the load capacitance. In the high-to-low transition, the NMOS already has a sufficient drain-to-source voltage from the load capacitance charge, so it can discharge the output without consuming any power from the supply. To summarize, in an output which switches from low to high and back to low again, the PMOS initially turns on, taking $CV_{DD}^2$ energy from the supply, then the NMOS will turn on, discharging $\frac{1}{2}CV_{DD}^2$ from the charged load capacitor (the other $\frac{1}{2}CV_{DD}^2$ is dissipated as heat in the resistive load).

\section*{Supplementary Note 5: Resetting a `receiverless' circuit}

\begin{figure}
    \centering
    \includegraphics[scale=0.6]{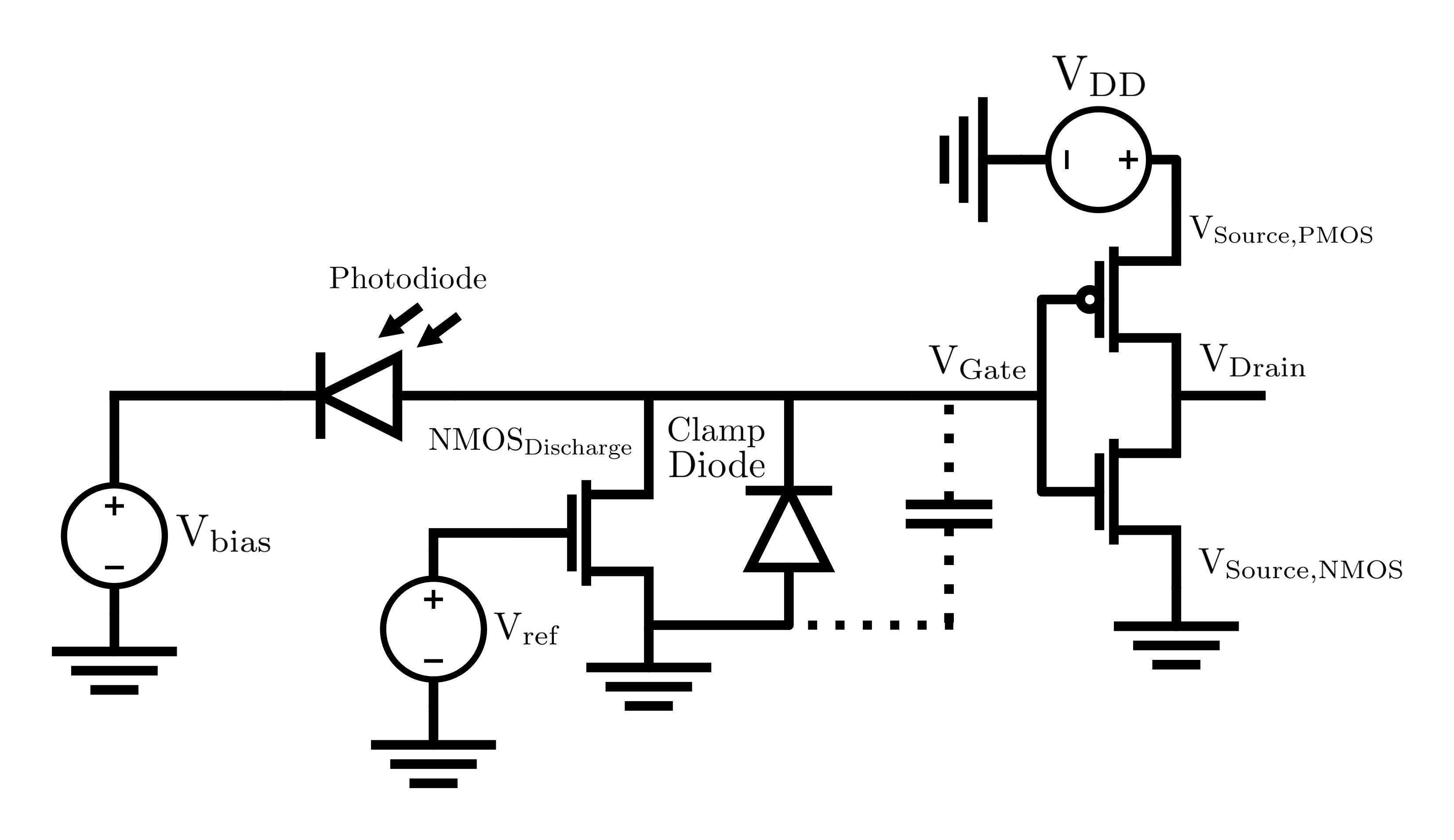}
    \caption{A proposed circuit for resetting the receiver lumped capacitor model.}
    \label{fig:reset_circuit}
\end{figure}

There are several circuit methods by which the accumulated charge on the input capacitor can be reset. In the method shown in Fig. \ref{fig:reset_circuit}, we place the NMOS device $\rm{NMOS}_{Discharge}$ between the photodetector and ground and drive the gate with an external reference voltage $V_\text{ref}$. The benefit of this solution is that it consumes no dynamic energy when there is no optical input power. However, it has the tradeoff that it requires additional area on chip and, because it is ratioed logic, requires careful design to ensure functionality. The width of $\rm{NMOS}_{Discharge}$ is set such that the accumulated charge on the capacitor generates a voltage high enough to overcome the input threshold of the load (modeled here as a CMOS inverter), but not so small that it cannot dissipate the charge quickly in a single clock cycle. One problem that arises from receiverless photodetection is that a constant steam of `1's coming into a photodetector without a strong enough $\rm{NMOS}_{Discharge}$ fill causes additional charge to slowly build on the load capacitance. To compensate, we propose a P-N junction diode ($\rm{Clamp~ Diode}$).

\section*{Supplementary Note 6: Electronic Repeaters}
A naive implementation of a repeater is a double inverter. The energy required is $C_\text{T} V_{\text{DD}}^2$, since in any transition, one inverter must be making a low-to-high transition and the other a high-to-low transition. As a result, in any `flip' of a repeater, one inverter does not consume energy. Using the values in Table~2 of the main text, the cost of a repeater is .06~fJ/bit for an output low-to-high transition. Therefore, even in the worst-case scenario where we place a repeater between every multiplier in an array of abutted 8-bit MAC units, the inter-multiplier interconnect energy cost is larger than that of the repeater.

\section*{Supplementary Note 7: Shot and thermal noise}
\begin{figure}[htbp]
    \centering
    \includegraphics[scale=.9]{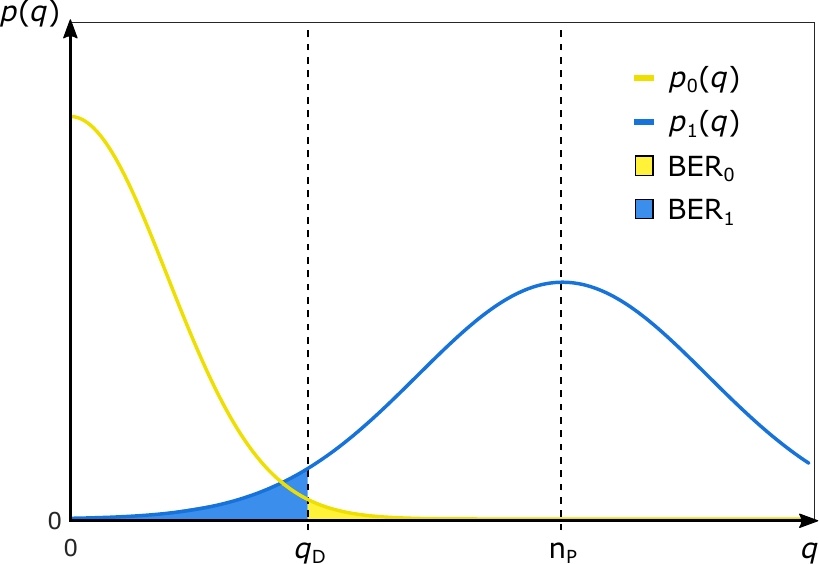}
    \caption{Schematic representation of probability density function of received charge (curves) and bit error rate (shaded region) - not to scale.}
    \label{fig:BER}
\end{figure}

In a hypothetical crosstalk-free DONN, the remaining noise sources are thermal (Johnson) and shot noise. To gain insight into whether they would affect classification accuracy, we estimate the ensuing bit error rates (BERs). The detector registers a `1' when $q \geq q_\text{D}$ photoelectrons are generated, and a `0' when $q<q_\text{D}$, where we assume the threshold charge is set by $q_\text{D} = n_\text{p}/2$ electrons. Fig.~\ref{fig:BER} illustrates the probability distributions of the number of photoelectrons, as well as the probabilities that a `0' is received when a `1' is sent (BER$_1$), and vice-versa (BER$_0$).

In a receiverless photodetector scheme, thermal noise can be approximated as `kT/C' noise~\cite{miller_attojoule_2017}, with:
\begin{align}
\sigma_{\text{V}}=\sqrt{k_BT/(C_\text{det}+C_\text{T})}
\end{align}

\noindent where $\sigma_\text{V}$ is the standard deviation of voltage, $k_B$ is the Boltzmann constant, $T$ is the temperature in Kelvin, $C_{\text{det}}$ is the capacitance of the photodetector, and $C_\text{T}$ is the capacitance of the inverter. The temperature depends on quality of heat sinking and proximity to hot spots; from Ref.~\cite{heat_2012}, we assume it is in the range $T\in\left[300-500\right]$. Using the values from Table 2 of the main text, we find $\sigma_\text{V}\approx5-6~\text{mV}\ll V_{DD}$. We can further verify whether thermal noise is likely to cause bit errors by approximating the probability distribution due to thermal noise, $p_\text{J}(q)$, by a Gaussian:
\begin{align}
        p_\text{J}(q) = \frac{1}{\sigma_\text{J}\sqrt{2\pi}}e^{-\tfrac{q^2}{2\sigma_\text{J}^2}}
\end{align}

\noindent with $\sigma_{\text{J}}=\sqrt{k_BT(C_\text{det}+C_\text{T})}/e \approx 6-7$~electrons.

\noindent To first order, shot noise will not affect the transmission of `0's (BER$_0$) since the number of transmitted photons is $n_\text{p}=0$. Thus:
\begin{align}
    \text{BER}_0
    = \sum_{q=q_D}^{\infty}p_0(q) = \sum_{q=q_D}^{\infty}p_\text{J}(q)
    &= \sum_{q=q_D}^{\infty}
    \frac{1}{\sigma_\text{J}\sqrt{2\pi}}e^{-\tfrac{q^2}{2\sigma_\text{J}^2}} \\
    &\approx \frac{1}{2}\text{erfc}\left( \frac{q_D}{\sqrt{2}\sigma_J} \right)
\end{align}

\noindent BER$_0$ for different $n_\text{p} = 2q_\text{D}$ are reported in Table~\ref{tab:BER}.

We assume shot noise follows a Poissonian probability distribution:
\begin{align}
    p_\text{shot}(q) =
    \frac{e^{-n_\text{p}}\left(n_\text{p}\right)^q}{q!}
\end{align}
\noindent where $n_\text{p}$ is the number of photons per detector per clock cycle.

\noindent For ease of computation with large $n_\text{p}$, we take the natural logarithm:
\begin{align}
    \text{ln}\left(p_\text{shot}(q)\right) &= \text{ln}\left(\frac{e^{-n_\text{p}}\left(n_\text{p}\right)^q}{q!}\right) \\
    &= \text{ln}\left(e^{-n_\text{p}}\right)+q\text{ln}\left(n_\text{p}\right)-\text{ln}\left(q!\right)\\
    &= -n_\text{p}+q\text{ln}\left(n_\text{p}\right)-\sum_{m=1}^q \text{ln}\left(m\right) \\
    &\Downarrow \\
    p_\text{shot}(q) &= \text{exp}\left(-n_\text{p}+q\text{ln}\left(n_\text{p}\right)-\sum_{m=1}^q \text{ln}\left(m\right)\right)
\end{align}

\noindent BER$_1$ due to shot noise is therefore:

\begin{align}
    \text{BER}_1^\text{shot} = \sum_{q=1}^{q_\text{D}-1}p_\text{shot}(q) \label{eq:BER1}
\end{align}

\noindent Results of this computation for various $n_\text{p}$ are shown in Table~\ref{tab:BER}.
\begin{table}[ht]
    \centering \caption{Expected values for BER$_1$ due to shot noise for different numbers of transmitted photons/bit}
    \begin{tabular}{c|c|c|c}
        $n_\text{p}$ & BER$_0^*$ & BER$_1^\text{shot}$ & BER$_1^\text{total}$ \\
        \hline
        10 & $10^{-1}$ & $ 10^{-2}$ & $10^{-1}$ \\
        100 & $10^{-18}-10^{-12}$ & $10^{-8}$ & $10^{-6}-10^{-5}$ \\
        1000 & small$^\dagger$ & $10^{-69}$ & $10^{-65}-10^{-63}$
    \end{tabular}
    \label{tab:BER}
    \caption*{*BER$_0=\text{BER}_1^\text{thermal}$ \\
    $^\dagger$Too small for MATLAB to compute. \\
    Note: We report a range since thermal noise, and therefore BER, depends on quality of heat sinking.}
\end{table}

Thermal noise will also contribute to BER$_1$; we convolve the probability distributions to find the total bit error rate:
\begin{align}
    \text{BER}_1^\text{total} = \sum_{q=1}^{q_\text{D}-1}p_1(q) = \sum_{q=1}^{q_\text{D}-1}p_\text{shot}(q)\circledast p_\text{J}(q)
\end{align}

From equation~(5) in the main text, we find $n_\text{p}\approx 1000$~photons/bit to generate a voltage swing of 0.8~V on the load capacitance; therefore, the expected BER is negligible, per Table~\ref{tab:BER}.

\bibliography{supplement}